\documentclass[10pt,conference]{IEEEtran}
\IEEEoverridecommandlockouts
\usepackage{amsmath,amssymb,amsfonts}
\usepackage{algorithmic}
\usepackage{graphicx}
\usepackage{textcomp}
\usepackage{xcolor}
\usepackage{hyperref}

\usepackage{silence}
\usepackage[
    style=ieee,
    giveninits=true,
    maxbibnames=6,
    minbibnames=1,
    doi=false,
    isbn=false,
    url=false,
    eprint=false
]{biblatex}
\usepackage{caption}
\usepackage{xspace}
\usepackage{makecell}
\usepackage{multirow}
\usepackage{tikz}
\usetikzlibrary{arrows.meta, positioning, fit, backgrounds}
\usepackage{booktabs}
\usepackage{soul}

\definecolor{midyellow}{HTML}{FFE599}    
\definecolor{midgreen}{HTML}{B6D7A8}     
\definecolor{midblue}{HTML}{A4C2F4}      
\definecolor{midlavender}{HTML}{B4A7D6}  
\definecolor{boldyellow}{HTML}{FFF200}  
\definecolor{boldgreen}{HTML}{00FF66}   

\newcommand{\DatasetName}{CloudIR-Bench\xspace}
\newcommand{\DatasetURL}{https://github.com/hoangtnm/CloudIR-Bench}
\newcommand{\RQOne}{Performance of LLMs on RCA\xspace}
\newcommand{\RQTwo}{Performance of LLMs on Mitigation Prediction\xspace}
\newcommand{\RQThree}{Performance of In-Context Learning with LLMs\xspace}
\newcommand{\RQFour}{Qualitative Analysis of LLM Failure Modes\xspace}

\def\BibTeX{{\rm B\kern-.05em{\sc i\kern-.025em b}\kern-.08em
    T\kern-.1667em\lower.7ex\hbox{E}\kern-.125emX}}

\usepackage{lineno}

\begin{document}


\title{\DatasetName: A Benchmark for Root Cause Analysis and Mitigation of Cloud Incidents}

\makeatletter
\newcommand{\linebreakand}{%
  \end{@IEEEauthorhalign}
  \hfill\mbox{}\par
  \mbox{}\hfill\begin{@IEEEauthorhalign}
}
\makeatother

\author{\IEEEauthorblockN{1\textsuperscript{st} Hoang Tran}
\IEEEauthorblockA{\textit{RMIT University} \\
Ho Chi Minh City, Vietnam \\
s4075740@student.rmit.edu.au}
\and
\IEEEauthorblockN{2\textsuperscript{nd} Huong Ha}
\IEEEauthorblockA{\textit{RMIT University} \\
Melbourne, Australia \\
huong.ha@rmit.edu.au}
\and
\IEEEauthorblockN{3\textsuperscript{rd} Tran Tri Dang}
\IEEEauthorblockA{\textit{RMIT University} \\
Ho Chi Minh City, Vietnam \\
tri.dangtran@rmit.edu.vn}
\linebreakand
\IEEEauthorblockN{4\textsuperscript{th} Jeffrey Chan}
\IEEEauthorblockA{\textit{RMIT University} \\
Melbourne, Australia \\
jeffrey.chan@rmit.edu.au}
\and
\IEEEauthorblockN{5\textsuperscript{th} Hongyu Zhang}
\IEEEauthorblockA{\textit{Chongqing University} \\
Chongqing, China \\
hyzhang@cqu.edu.cn}
}

\maketitle

\begin{abstract}
    Root cause analysis (RCA) and mitigation prediction are essential for incident management, enabling the identification of underlying causes of failures and ensuring the reliability of complex cloud-based systems. With rising incident volume and system complexity, incident reports have become a valuable resource for these tasks. Recent studies have shown promising results using Large Language Models (LLMs) to predict root causes and mitigation actions from textual descriptions in incident reports. Despite these advances, further progress is hindered by the lack of publicly available datasets; existing work relies on private data, which limits transparency, reproducibility, and generalization. To address this gap, we introduce \DatasetName, the first publicly available benchmark providing a dataset and evaluation framework for incident report-based RCA and mitigation prediction.
    Curated from incident reports across five leading cloud-based systems, \DatasetName captures a diverse range of real-world production scenarios representing authentic failures and engineers' reasoning. We also release an open-source evaluation framework that includes eight recent LLM-based approaches. Our comprehensive evaluation includes zero-shot and in-context learning experiments, along with a qualitative analysis to identify where and why LLMs fail. Benchmarking results reveal the capabilities and limitations of existing LLMs in this domain. The benchmark, including the dataset, framework and supplementary materials, is available at \href{\DatasetURL}{\DatasetURL}.
\end{abstract}

\begin{IEEEkeywords}
Incident Management, Incident Report, Root Cause Analysis, Cloud Services, Large Language Models
\end{IEEEkeywords}

\section{Introduction} 

Over the past decade, advances in cloud computing technologies have driven a widespread migration of software systems from on-premises servers to cloud platforms such as Amazon Web Services (AWS), Microsoft Azure, and Google Cloud. Despite significant investments in cloud system reliability~\cite{ahmed2023recommending,zhang2024automated,chen2024automatic,goel2024x}, unplanned incidents (e.g., service failures, unexpected disruptions, and performance degradations) remain unavoidable. These incidents can severely impact organizational profitability and customer satisfaction. For example, a single hour of downtime during peak sales 
costs Amazon up to \$100 million in revenue loss \cite{wolfe2018amazon}.

When an incident occurs, it is essential to quickly identify the root cause and propose an effective mitigation strategy to minimize operational and financial impact. As such, root cause analysis (RCA) and mitigation prediction have become crucial to ensure the reliability of modern cloud-based systems. While traditional RCA research primarily leverages telemetry data (logs, metrics, and traces) \cite{pham2024baro, zheng2024mulan, zhang2025failure, pham2026torai}, recent advances in Large Language Models (LLMs) have enabled a new research direction: performing RCA directly from unstructured incident reports \cite{ahmed2023recommending,chen2024automatic,zhang2024automated, goel2024x, saha2022mining,roy2024exploring}. These reports capture rich contextual information, including service interactions, human-level reasoning, and customer impacts, which are often invisible in raw telemetry data. Consequently, incident reports have become a key resource for effective RCA and mitigation planning, supporting both rapid incident response and long-term system reliability. For example, the work in \cite{ahmed2023recommending} was the first to leverage LLMs trained on a private historical dataset of incident reports, where each report includes details on incident symptoms, root causes, and mitigation strategies, to identify root causes and mitigation plans for new incidents. 
The work in \cite{zhang2024automated} proposed an in-context learning approach enabling LLMs to learn effectively from historical incident reports and improve RCA and mitigation prediction for new incidents. 
Furthermore, the work in~\cite{liu2023ticket} demonstrates that fine-tuning language models on incident ticket titles and descriptions can achieve highly accurate multi-class classification of incident tickets, complementing free-text root cause prediction by enabling rapid incident triage and automated routing to appropriate resolution teams in production systems.

Despite the promising results of LLM-based approaches for incident report-based RCA reported in these studies, this research direction still faces significant challenges:
\begin{itemize}
    \item \textbf{Lack of public datasets:} Existing work relies exclusively on private data, which is not publicly shared due to confidentiality and privacy constraints. This lack of accessibility limits reproducibility, hinders standardized evaluation, and impedes broader research progress. Currently, there is no publicly available dataset or benchmark for RCA and mitigation prediction based on incident reports.
    \item \textbf{Challenges in dataset construction:} Building a well-curated dataset of incident reports is technically challenging. Incident reports often vary significantly in format and language, requiring extensive preprocessing. Moreover, extracting root causes and mitigation actions typically requires manual expert annotation, as such information is often implicit or loosely defined within each report.
\end{itemize}


To address these challenges and advance research on incident report-based RCA and mitigation prediction, we introduce \textbf{\DatasetName}, the first publicly available benchmark that includes both a dataset and an evaluation framework for these tasks in cloud-based systems. Our dataset contains incident reports curated from five leading cloud-based systems, covering a wide range of real-world production incidents, including configuration errors, software/hardware failures, network issues, cascading failures, and service dependency failures. These reports were originally documented by the engineers responsible for handling the incidents, and we extract the detailed incident information, along with the corresponding root causes and mitigation actions as ground-truth labels for the RCA and mitigation prediction tasks. Specifically, each dataset entry contains the title, summary, impact, timeline, along with the corresponding root cause and mitigation plans, all presented in free-form text. As such, \DatasetName serves as a rich textual resource for AI-driven RCA and mitigation prediction, supporting not only these key tasks, but also related tasks such as information retrieval and processing of unstructured incident reports. For evaluation, \DatasetName includes an open-source framework that provides a unified experimental environment for evaluating recent LLMs in RCA and mitigation prediction. The framework leverages standard APIs for LLM inference, simplifying integration and enabling seamless model interchangeability. We further evaluate recent LLMs to reveal key insights into their capabilities for identifying root causes and predicting mitigation plans from incident reports, motivating further research in this area.

In summary, our key contributions are as follows:

\begin{itemize}
    \item \textbf{The first public incident report dataset}: We introduce \DatasetName, the first publicly available dataset of incident reports curated from five leading cloud-based systems. The dataset is enriched with long-context information and metadata to support a deep understanding of incident diagnosis and mitigation. It is constructed following established best practices to ensure data quality and promote transparency in RCA and mitigation prediction research. It is also highly diverse, representing a wide range of incident characteristics.  
    \item \textbf{An evaluation framework for incident report-based RCA and mitigation prediction}: \DatasetName includes an open-source evaluation framework that enables fair comparisons of recent LLM baselines for incident report–based RCA and mitigation prediction using standard APIs for LLM inference.
    \item \textbf{A comprehensive evaluation of recent LLMs}: Using the proposed dataset and evaluation framework, we conduct an extensive evaluation and analysis of recent LLMs on RCA and mitigation prediction, providing critical insights into the LLMs' capabilities and limitations.
\end{itemize}

\section{Background}

\subsection{Incident Management}

Production incidents remain a persistent challenge in large-scale cloud-based systems, negatively affecting customer satisfaction and requiring substantial costs for RCA and mitigation efforts.
A standard incident lifecycle typically includes four stages: (1) Detection, which involves the identification of anomalous system behaviors, through automated monitoring services and customer reports \cite{hadadi2024anomaly,mahindru2021log}; (2) Triaging, which involves engineers to assess the severity of reported anomalies and assign them to the appropriate team for initial investigation~\cite{pham2020deeptriage,bansal2020decaf}; (3) Diagnosis, during which assigned teams engage in iterative communication and analysis to determine the nature of the incident and identify its root cause \cite{ma2020diagnosing,bansal2020decaf}; (4) Mitigation, where engineers conduct mitigation strategies based on the identified root cause to resolve the issue and restore the system to normal operation \cite{ahmed2023recommending,jiang2020mitigate}.

\subsection{RCA and Mitigation Prediction for Incidents}
RCA and mitigation prediction are critical stages of incident management, enabling engineers to identify the underlying causes of failures, implement effective mitigation strategies, and prevent future recurrence \cite{chen2020towards}. However, performing RCA and mitigation prediction in cloud environments is complex and resource-intensive, requiring not only expert knowledge but also the ability to interpret heterogeneous and unstructured system data \cite{ma2020diagnosing}. As a result, these processes are not only time-consuming but also error-prone, further complicating the incident management process.

Traditional RCA research primarily relies on \textit{telemetry-based} approaches using logs, metrics, and traces \cite{pham2024causalRCA, hadadi2024anomaly, ganatra2023detection, zhao2023robust}. While effective for localized fault detection, these approaches often fail to capture full incident context, including service interactions, human reasoning, and customer impacts, which are essential for understanding complex failures.


Our work instead focuses on \textit{incident report-based} RCA, which is a recent and increasingly important research direction. Given unstructured textual descriptions of incident reports, the goal is to generate the root cause and recommend mitigation strategies in free-form text~\cite{ahmed2023recommending,zhang2024automated,chen2024automatic,jiang2020mitigate}. Incident reports, written by engineers during and after incident resolution, capture rich contextual information including system behaviors, cross-service dependencies, human observations, and mitigation actions often invisible in telemetry data. This represents a fundamentally different input modality and research problem from telemetry-based RCA. By analyzing historical incident reports, engineers can gain a deeper understanding of the incident lifecycle, service interactions, and effective mitigation strategies. Moreover, historical incident reports reveal recurring patterns, support predictive modeling, and provide valuable training data for automated RCA and mitigation prediction. For example, BugFixer~\cite{hu2014effective} and DeepRmd \cite{jiang2020mitigate} demonstrate the feasibility of recommending responsible engineers or retrieving relevant troubleshooting documents that describe mitigation processes from historical incident reports. 

\subsection{LLMs for RCA and Mitigation Prediction}
Recently, the lifecycle of incident management has been transformed by the adoption of LLMs \cite{ahmed2023recommending,chen2024automatic,zhang2024automated,saha2022mining} to enable more sophisticated RCA and mitigation prediction from historical incident reports. \citeauthor{ahmed2023recommending} \cite{ahmed2023recommending} was the first to apply fine-tuned LLMs, specifically GPT-3.x variants, to cloud incident reports to assist engineers in RCA and mitigation prediction. Fine-tuned on a private dataset of incidents within Microsoft, the study compared multiple training settings and evaluated outputs using both semantic and lexical metrics. The results demonstrated that LLMs can accurately identify the underlying root causes and predict mitigation actions, even in cases where such information is challenging for human experts to identify. Building on this work, subsequent research has proposed to apply advanced prompting strategies, including in-context learning (ICL) and chain-of-thought (CoT) reasoning, to further improve performance on RCA and mitigation prediction from incident reports~\cite{zhang2024automated, chen2024automatic, roy2024exploring,goel2024x}.

It is also worth noting that, in general text generation tasks, a notable performance gap exists between closed-source commercial models (e.g., GPT-5.2, GPT-4.1) and open-weight alternatives such as Llama 3~\cite{dubey2024llama}, Mixtral \cite{jiang2024mixtral}, and DeepSeek-R1 \cite{guo2025deepseek}. Understanding how this performance gap affects the RCA and mitigation prediction tasks remains unknown.

\subsection{Datasets for RCA and Mitigation Prediction}
Despite the demonstrated success of LLMs for RCA and mitigation prediction, progress in incident report-based RCA research is hindered by the lack of publicly available datasets, as existing datasets are private \cite{ahmed2023recommending}, limiting reproducibility and transparency. While telemetry-based RCA datasets (e.g., LEMMA-RCA \cite{zheng2024lemma}, OpenRCA \cite{xu2025openrca}, RCAEval \cite{pham2025rcaeval}) exist, they address a fundamentally different problem from incident report-based RCA. These telemetry-based benchmarks cannot be adapted for incident report analysis as they lack the unstructured textual context, human reasoning, and cross-service insights captured in engineer-written incident reports.

Recognizing the critical need for a public benchmark to advance incident report-based RCA and mitigation prediction research, this work introduces the first publicly available dataset of authentic production incident reports. Our dataset includes diverse types of incident reports from five leading cloud-based systems, capturing real-world engineer reasoning, broader contextual information, and realistic RCA challenges. Additionally, we develop an open-source evaluation framework that enables fair comparisons of recent LLM baselines for incident report–based RCA and mitigation prediction
using standard APIs for LLM inference. Finally, we conduct a comprehensive analysis of recent LLMs to provide insights into their current capabilities and limitations in RCA and mitigation prediction in cloud incidents.

\section{The \DatasetName Dataset}

\subsection{Objectives} \label{dataset-objectives}

We aim to establish a comprehensive dataset of real-world incident reports from leading cloud-based systems. The goal is to provide a standard benchmark for research on incident report-based RCA and mitigation prediction. Each report in the dataset includes detailed fields, such as title, summary, impact, root cause, mitigation, and timeline, facilitating a range of downstream research tasks. The construction of the dataset is driven by the following criteria:

\begin{enumerate}
    \item \textbf{Real-world relevance:} we prioritize large-scale, actively maintained cloud-based systems with a substantial number of global users and long development histories. This ensures the dataset reflects the true complexity of modern production environments. 
    \item \textbf{Diversity:} each chosen system’s incident database must cover a wide range of services (e.g., compute, storage, networking) and failure types, ranging from software issues and configuration errors to hardware malfunctions and cascading service dependencies.
    \item \textbf{Reproducibility and automatability:} A high level of automation is essential for creating a public, reproducible dataset by minimizing human errors and biases. This requires selecting cloud providers with robust development and annotation processes adhering to industry standards. To further support reproducibility, we open-source our dataset construction code, allowing others to validate, replicate, and build upon our work.
\end{enumerate}

\begin{figure*} [ht]
    \centering
    \includegraphics[width=0.99\linewidth]{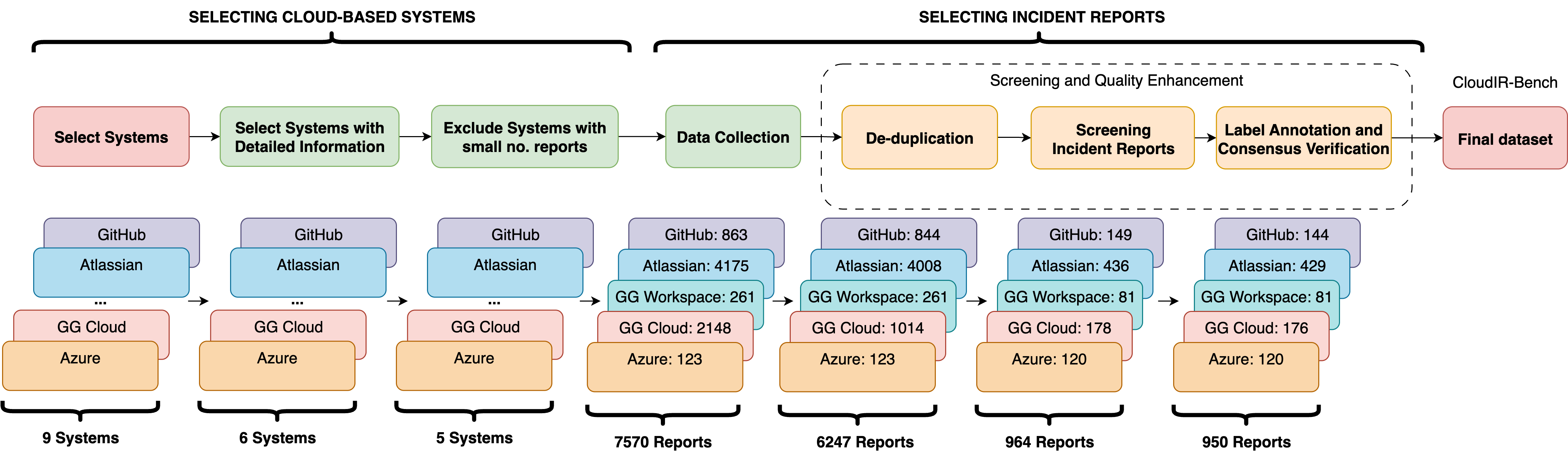}
    \caption{Methodology for dataset construction.}
    \label{fig:data-construction}
\end{figure*}

Figure \ref{fig:data-construction} illustrates the dataset construction workflow. It first identifies suitable cloud-based systems and then extracts relevant information from their incident reports (e.g., title, summary, root cause, and mitigation steps) through a pipeline that combines automated extraction and manual review.

\subsection{Selecting Cloud-based Systems}

Guided by the aforementioned criteria, we identified nine cloud-based systems as potential candidates for inclusion in the dataset: Microsoft Azure \cite{azure}, Google Cloud \cite{google_cloud}, Google Workspace \cite{google_workspace}, Atlassian \cite{atlassian}, AWS~\cite{aws}, GitHub~\cite{github}, CircleCI \cite{circleci}, Notion \cite{notion}, and Alibaba Cloud \cite{alibaba_cloud}. These systems publish post-incident reviews, documented by the engineers in charge, that provide detailed descriptions of incidents after resolution, including incident summaries, root causes, mitigation plans, and other relevant information. Note that, for the incident summaries, our goal is to construct a dataset of incident reports based on the information available at the time of initial incident reporting, so from the post-incident summaries, we keep only the details presented during the initial incident reporting.


We then conducted multiple rounds of manual review for each candidate system, carefully examining their incident reports to assess whether they include detailed incident summaries, root causes, and mitigation steps. During this process, CircleCI, Notion, and Alibaba Cloud were excluded due to the lack of sufficient detail in root cause and mitigation descriptions in most of their incident reports. Among the remaining six systems, AWS was also excluded because it provided a limited number of incident reports (only 17), despite its high quality and thorough documentation.

Finally, we selected five cloud-based systems, including Microsoft Azure, Google Cloud, Google Workspace, Atlassian, and GitHub, to construct the dataset, ensuring both depth and breadth of incident coverage across diverse operational contexts. This selection process aims to balance comprehensiveness with quality and representativeness of the dataset.

\subsection{Data Collection}

The first step involves crawling raw incident reports from each target data source. Since none of these sources provides official APIs for automated data collection, we employed web scraping techniques as an alternative. Specifically, we developed custom web crawling techniques using Selenium and Beautiful Soup to extract the required information. By parsing and navigating the HTML layouts, we ensure robust and precise data extraction. 
To accommodate the distinct layouts of each platform, we implemented separate scraping scripts, each tailored to the unique HTML structure of its respective data source. The resulting implementation comprises approximately 1,500 lines of code, enabling us to unify the extracted data into a standardized and consistent JSON format.


\subsection{Screening and Quality Enhancement}

Our analysis of the collected data from the five selected cloud-based systems reveals that a substantial portion of the raw incident reports lacks meaningful content and includes duplicate incident reports. Specifically, in many cases, the reports consist of only brief summaries stating that the issue has been resolved while omitting critical information such as the root cause or mitigation steps. As a result, these reports offer limited value for the dataset. To address this issue, we first removed duplicates and kept only reports with substantive information such as title, summary, root cause, and mitigation steps (Section \ref{sec:dedup}). We then excluded any reports that lack either the root cause or mitigation steps, as they do not provide useful insights (Section \ref{sec:screen-ir}). Furthermore, since each organization uses a distinct format for documenting incidents, we develop customized filtering and extraction strategies for each system to ensure consistent and relevant information is extracted from the incident reports (Section \ref{sec:label-anno}).


\subsubsection{De-duplication} \label{sec:dedup}

In large-scale cloud-based systems, many services and components are often tightly interconnected. As a result, a single underlying fault can propagate across multiple services, leading to simultaneous failures and causing the same incident to be reported multiple times across different services/components. This necessitates a de-duplication process to ensure the uniqueness of incident reports. Since each system employs different mechanisms for tracking and documenting incidents, we design tailored de-duplication strategies for each system to accurately identify and remove redundant ones, ensuring each incident is represented only once in the dataset. For example, when multiple incident reports share the same incident ID within a system, we keep only the earliest report and discard the duplicates.

A detailed description of these de-duplication methods is provided in Section A1 in the appendix \cite{appendixlink}. 

\subsubsection{Screening Incident Reports} \label{sec:screen-ir}

In this step, we screen the collected incident reports obtained from the previous step to ensure that the dataset includes only those reports with valid root cause and mitigation information. Moreover, given the large volume of reports in this collection, manually labeling all of them would be inefficient and error-prone. To address this, we first apply an automated keyword-based filtering method using predefined extraction patterns (detailed in Section A2 in the appendix \cite{appendixlink}) to exclude reports that are unlikely to contain relevant root cause or mitigation information, thereby reducing the dataset size. We then manually review the filtered reports to verify their relevance and to minimize the risk of excluding valuable reports.

\subsubsection{Label Annotation and Consensus Verification} \label{sec:label-anno}
\hfill \\
\textbf{\\Label Annotation.} We extracted key attributes from each incident report, including Title, Summary, Impact, Root Cause, Mitigation, Updates, Severity, Status, and Timestamp. While structured fields (severity, status, timestamp) can be extracted automatically, most attributes require manual annotation due to their unstructured nature.
To preserve the authenticity of the original public reports, we keep the extracted fields in their original form rather than paraphrasing them. Special emphasis is placed on the fields “Summary”, “Root Cause”, and “Mitigation” as they are critical for understanding the system failures, identifying the root causes, and predicting the mitigation strategies. Furthermore, given that our data sources are post-incident reviews, some summaries may include sentences that might reveal root causes or mitigation. We call this ``information leakage". To address this, we identify and remove such information from incident reports, ensuring that root-cause or mitigation information is not accessible to LLMs during evaluation. To reduce bias, each incident report is independently annotated by two authors with software engineering expertise, followed by a consensus verification.

\textbf{Consensus Verification.} 
Since each report is annotated by two labelers, disagreement may arise. These are resolved through discussion. Furthermore, to understand the inter-annotator agreement across the labelers, we compute the Cohen's Kappa score $\kappa$. The resulting scores indicate strong overall agreement. Specifically, Summary, Root Cause, and Mitigation achieve $\kappa$ of 0.89-0.91 for Azure, Google Cloud, and Google Workspace, and 0.63-0.64 for GitHub and Atlassian, while temporal fields such as Start-time, End-time, and Updates achieve $\kappa$ of 0.71-1.00 across all datasets. These results indicate strong agreement among the labelers.

Details of these two steps are provided in Section A3 in the appendix.
Fig. \ref{fig:sample-incident} presents an example of a finalized incident report after completing the full labeling process.

\begin{figure}[ht]
\fbox{\begin{minipage}{0.98\columnwidth}\small
\textbf{Title:}
Incident affecting Gmail

\textbf{Summary:} 
On Wednesday, 27th April 2022, Gmail experienced delayed email delivery globally.

\textbf{Impact:}
On Wednesday, 27th April starting from 14:30 US/Pacific, Gmail delivery was unavailable or delayed for messages to and from some customers globally. Some customers experienced delays in receiving email. A subset of inbound emails in this period were rejected by our APIs with a request for client systems to retry them.

\textbf{Reference root cause:} 
Google engineers often make configuration changes to optimize system performance. Recently, one such change adversely affected how Gmail manages traffic. As a result, some customers experienced delays in receiving email. A subset of inbound emails in this period were also rejected by our APIs with a request for client systems to retry them

\textbf{Reference mitigation:}
Google engineers rolled back the configuration change, which prevented any further impact. Services were restored for all users within 1 hour and 5 minutes.

\textbf{Long-term mitigation:}
*Re-examining and enhancing the process through which configuration changes are rolled out.
*Examining and improving how traffic is routed through our internal systems when configuration changes are made.
*Rolling out future changes in stages instead of globally.

\textbf{Updates:}
22:32 UTC on 27 Apr 2022: We're investigating reports of an issue with Gmail. We will provide more information shortly.
22:58 UTC on 27 Apr 2022: The problem with Gmail has been resolved. Customers may have experienced delays with email delivery between 14:30-15:35 US/Pacific.


\textbf{Status:}
resolved

\textbf{Timestamp:}
Start: 2022-04-27 21:30 - End: 2022-04-27 22:35

\end{minipage}}
\centering

\caption{A sample incident report of Google Workspace.
}
\label{fig:sample-incident}
\end{figure}

\subsection{Statistics of the Final Dataset}


Table \ref{tab:data-stats-detailed} describes the statistics of final incident reports in our dataset. \DatasetName aggregates 950 authentic production incident reports from five leading large-scale cloud systems, with 911 root causes and 832 mitigation strategies. Among these systems, Atlassian contributes the most incident reports (429), followed by Google Cloud (176), GitHub (144), Azure (120), and Google Workspace (81).

\DatasetName covers an extensive range of production failures and fault types across diverse cloud infrastructures, representing real-world scenarios from multiple organizations rather than being restricted to a single company's internal data.
Failure categories include configuration errors, service dependencies, cascading failures, resource exhaustion, software bugs, and infrastructure issues. This multi-organizational scope gives \DatasetName broader fault coverage than single-organization proprietary datasets, making it comparable to or more representative than industry-private alternatives, while remaining publicly accessible for reproducible research.


\begin{table*}
  \centering
  \caption{Detailed Statistics of the Final Dataset}
  \label{tab:data-stats-detailed}
  \begin{tabular}{llccccc}
    \toprule
    \textbf{Attribute} & \textbf{Description} & \textbf{Azure} & \textbf{Google Cloud} & \textbf{Google Workspace} & \textbf{GitHub} & \textbf{Atlassian} \\
    \midrule
    Title                & Incident title & 120 & 176 & 81  & 144 & 429 \\
    Summary              & Incident summary & 120 & 176 & 81  & 144 & 428 \\
    Impact               & Reported impact & 0   & 175 & 81  & 0   & 155 \\
    Root cause           & Cause of the incident & 119 & 175 & 81  & 142 & 394 \\
    Mitigation           & Resolution actions & 119 & 148 & 76  & 134 & 355 \\
    Long-term mitigation & Preventive measures & 100 & 86  & 69  & 125 & 175 \\
    Updates              & Timeline of actions/notes & 20  & 176 & 81  & 144 & 424 \\
    Severity             & Incident severity & 120 & 176 & 81  & 144 & 429 \\
    Status               & Current status & 120 & 176 & 81  & 144 & 429 \\
    Start\_timezone      & Detection timezone & 120 & 176 & 81  & 144 & 429 \\
    Start\_time          & Detection time & 120 & 176 & 81  & 144 & 429 \\
    End\_timezone        & Resolution timezone & 120 & 176 & 81  & 144 & 429 \\
    End\_time            & Resolution time & 120 & 176 & 81  & 144 & 429 \\
    \bottomrule
  \end{tabular}
\end{table*}

\section{Evaluation Framework }

\begin{figure*}[ht]
    \centering
    \includegraphics[width=0.95\linewidth]{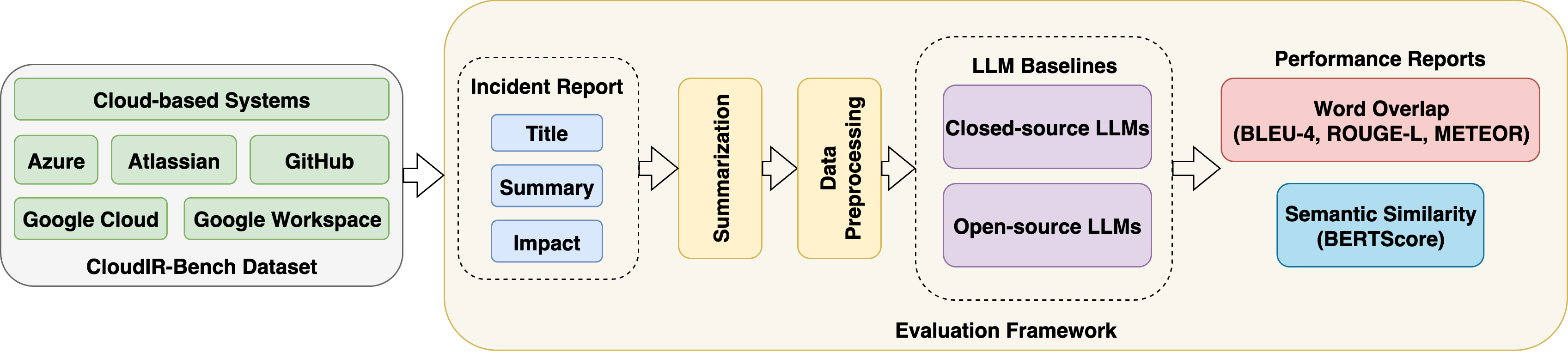}
    \caption{Methodology for evaluation.}
    \label{fig:eval-workflow}
\end{figure*}

To evaluate our proposed \DatasetName dataset and understand the current state of LLM-based methods for RCA and mitigation prediction, we provide an open-source evaluation framework alongside the dataset. Our evaluation framework includes 8 LLM baselines, covering a wide range of closed-source and open-weight LLMs, and uses standard APIs for LLM inference. It also supports data preprocessing and benchmarking with both lexical and semantic metrics. Detailed documentation is available on our GitHub repository \cite{appendixlink}, including instructions for installation, usage, and steps for integrating other baselines. The documentation also provides example workflows and guidance to ensure reproducibility.


To thoroughly assess model performance and understand the underlying factors that affect the prediction quality, we investigate the following research questions (RQs):
\begin{itemize}
    \item \textbf{RQ1:} How effective are LLM-based methods in identifying the root causes from incident reports?
    \item \textbf{RQ2:} How effective are LLM-based methods in predicting the mitigation plans from incident reports?
    \item \textbf{RQ3:} How effective is In-Context Learning in improving the performance of LLMs?
    \item \textbf{RQ4:} Where and why do LLMs fail in RCA and mitigation prediction tasks?
\end{itemize}



\subsection{Evaluation Baselines}

To evaluate the proposed benchmark dataset and gain insights into LLM-based RCA and mitigation prediction methods, we conduct experiments comparing baseline models from prior work with other recent LLMs.
For closed-source models, we select GPT-4.1 and GPT-5.2 as the direct successors to GPT-3.5 and GPT-4 variants used in prior work~\cite{ahmed2023recommending,zhang2024automated}, but are now deprecated.
For open-weight models, we select top-ranked and widely adopted LLMs on SWE-bench while covering various model sizes and architectures.
Figure \ref{fig:eval-workflow} illustrates the experimental workflow, outlining key stages including label generation, model inference, and evaluation. 

The evaluation begins by using the title, summary, and impact of an incident report as input to the LLM-based methods. Since incident reports are often lengthy and contain extensive details, they may introduce noise that obscures the core incident information and complicates downstream inference and evaluation tasks \cite{ahmed2023recommending, zhang2024automated, chen2024automatic}. To address this challenge, following prior works \cite{zhang2024automated,chen2024automatic}, we preprocess each incident report using an LLM-based summarization module, specifically OpenAI’s GPT-4.1, to generate a concise summary that captures the essential information of the report.
Since leaked sentences are removed prior to summarization, there is no exposure of root cause or mitigation beyond what was already filtered during annotation. Both the incident title and the summarized summary are then used as inputs to LLMs to predict root causes and mitigation.

We evaluate the following closed-source and open-weight LLMs:
\begin{itemize}
    \item \textbf{GPT-5.2} and \textbf{GPT-4.1}: OpenAI's SOTA models, with exceptional coding abilities and instruction following.

    \item \textbf{DeepSeek-R1-Distill-Qwen-32B} and \textbf{DeepSeek-R1-Distill-Llama-8B} \cite{guo2025deepseek}: Distilled, efficient, instruction-tuned variants of DeepSeek-R1.
    \item \textbf{Qwen3-32B} and \textbf{Qwen3-30B-A3B} \cite{qwen3}: 30B+ parameter LLMs and a 3B-active Mixture-of-Experts variant, with long context and strong reasoning.
    \item \textbf{Llama-3.1-8B-Instruct \cite{dubey2024llama}}: An enhanced version of Meta’s Llama 3, optimized for improved performance and better instruction following.
    \item \textbf{Ministral-8B-Instruct-2410 \cite{Ministral8BInstruct2410}}: Instruction-tuned model from Mistral AI, featuring strong reasoning abilities and support for long context lengths.
\end{itemize}

\subsection{Evaluation Metrics}
To quantitatively assess model performance on \DatasetName, we compare predictions against the ground-truth root causes and mitigation actions using both lexical and semantic metrics, following established evaluation protocols in prior work \cite{ahmed2023recommending,zhang2024automated,roy2024exploring,saha2022mining,goel2024x,wang2024rcagent}. While RCA and mitigation prediction involve semantic reasoning, these metrics capture meaningful signals. Lexical overlap reflects whether key technical terms (e.g., service names, failure types) are correctly identified, while semantic similarity measures whether the predictions have similar meaning even when phrased differently.
We do not adopt LLM-as-a-judge in this work due to three limitations: (1) LLMs tend to assign higher scores when used for both prediction and evaluation \cite{zheng2023judging}; (2) accurate RCA requires domain-specific knowledge beyond general-purpose LLMs; and (3) non-determinism and prompt sensitivity lead to inconsistent scores across inference runs.


\subsubsection{Lexical Metrics}
Following existing work for RCA \cite{ahmed2023recommending,zhang2024automated,goel2024x,roy2024exploring}, we employ three standard lexical metrics. We adopt \textbf{BLEU-4} \cite{papineni2002bleu}, which quantifies $n$-gram overlap from unigrams to 4-grams between model predictions and ground-truth labels. We then choose \textbf{ROUGE-L} \cite{lin2004rouge} to assess sentence-level structural similarity by identifying the longest matching subsequence between the prediction and reference through Longest Common Subsequence (LCS). Lastly, we use \textbf{METEOR} \cite{banerjee2005meteor} to evaluate the harmonic mean of unigram precision and recall while incorporating features such as stemming and synonym matching.
For RCA and mitigation prediction, a higher score indicates that the prediction shares more technical entities with the label, such as affected services, configuration parameters, suggesting more accurate identification of the incident's key elements and vice versa.


\subsubsection{Semantic Metric}
Since root cause and mitigation strategies can be expressed in various ways, lexical metrics that rely on word-level matching may fail to capture the true semantic similarity  \cite{ahmed2023recommending,zhang2024automated,goel2024x,roy2024exploring}. To address this, we perform semantic evaluation alongside lexical evaluation to ensure a more comprehensive assessment that reflects the underlying meaning.
Specifically, we employ BERTScore \cite{zhang2019bertscore}, a widely used semantic evaluation metric in RCA research \cite{ahmed2023recommending,zhang2024automated,goel2024x,roy2024exploring}, which uses contextualized embeddings from the pretrained language model BERT to compute similarity scores based on token-level cosine similarity. This approach captures semantic equivalence in linguistics.

It's worth noting that while lexical and semantic metrics are informative and can measure textual overlap or similarity with the ground truth, they do not always capture causal correctness. A prediction can achieve a high similarity score while misidentifying the underlying root cause and actual mitigation steps. We examine this limitation in our qualitative analysis (Section~\ref{sec:rq4}) and discuss its implications in Section~\ref{sec:threats}.

\subsubsection{Top1 and Top5 Scores}

Following~\cite{ahmed2023recommending}, we compute Top1 and Top5 scores for each evaluation metric. For each incident, we generate ten predictions to capture output diversity. The Top1 score is computed by averaging the metric scores of the top-1-ranked prediction (based on the LLM's confidence scores) across all incidents. The Top5 score is computed by selecting the highest metric score among the top-5-ranked predictions for each incident and averaging these scores across all incidents. Top1 reflects the quality of the model’s most-confident prediction, whereas Top5 reflects the best achievable performance when selecting from the five highest-ranked predictions. Thus, Top5 represents an upper bound on the performance achievable by a downstream selection mechanism (e.g., human review of a shortlist) that can reliably identify the best candidate from the top-5-ranked predictions.


\subsection{Evaluation Results}

\subsubsection{RQ1: \RQOne} \label{sec:rq1}

\begin{table*}
\centering
\caption{RCA evaluation results. Best results are in bold.} 
\label{tab:results-root_cause}
\begin{tabular}{llcccccccc}
\toprule
\multirow{2}{*}{Model} & \multirow{2}{*}{Dataset} & \multicolumn{2}{c}{BLEU-4} & \multicolumn{2}{c}{ROUGE-L} & \multicolumn{2}{c}{METEOR} & \multicolumn{2}{c}{BERTScore} \\
& & Top1 & Top5 & Top1 & Top5 & Top1 & Top5 & Top1 & Top5 \\
\midrule
    GPT-5.2 & Azure & 1.57 & 0.55 & 13.88 & 16.72 & 19.05 & 23.36 & 80.78 & 81.77 \\
    GPT-4.1 & Azure & \textbf{2.25} & \textbf{1.74} & 14.86 & 19.48 & 19.19 & 25.23 & \textbf{81.66} & \textbf{82.92} \\
    DeepSeek-R1-Distill-Qwen-32B & Azure & 1.41 & 1.08 & 12.46 & 17.09 & 13.49 & 18.91 & 79.1 & 81.03 \\
    Qwen3-32B & Azure & 2.17 & 1.38 & \textbf{15} & 18.94 & 16.83 & 21.88 & 81.2 & 82.68 \\
    Qwen3-30B-A3B & Azure & 1.15 & 1.16 & 12.1 & 17.6 & 12.09 & 16.57 & 79.34 & 81.84 \\
    DeepSeek-R1-Distill-Llama-8B & Azure & 1.61 & 1.15 & 12.31 & 16.1 & \textbf{19.75} & \textbf{24.58} & 79.53 & 81 \\
    Llama-3.1-8B-Instruct & Azure & 2.09 & 1.45 & 13.47 & \textbf{19.2} & 17.33 & 24.34 & 80.27 & 82.12 \\
    Ministral-8B-Instruct-2410 & Azure & 0.69 & 0.63 & 9.61 & 16.71 & 9.4 & 18.21 & 77.45 & 81.08 \\ 
    \midrule
    
    GPT-5.2 & Google Cloud & 0.97 & 0.97 & 12.39 & 15.84 & 18.22 & 23.17 & 79.82 & 80.88 \\
    GPT-4.1 & Google Cloud & 1.93 & \textbf{1.61} & \textbf{14.41} & 18.46 & 19.74 & \textbf{26.18} & \textbf{81.02} & 82.14 \\
    DeepSeek-R1-Distill-Qwen-32B & Google Cloud & 1.85 & 0.86 & 12.74 & 18.18 & 15.48 & 21.88 & 79.52 & 81.42 \\
    Qwen3-32B & Google Cloud & 1.84 & 1.08 & 13.98 & 18.3 & \textbf{18.30} & 23.96 & 80.79 & \textbf{82.18} \\
    Qwen3-30B-A3B & Google Cloud & \textbf{2.00} & 0.71 & 12.57 & 17.77 & 14.25 & 18.65 & 80.11 & 81.92 \\
    DeepSeek-R1-Distill-Llama-8B & Google Cloud & 1.22 & 0.97 & 11.63 & 15.82 & 19.6 & 25.46 & 79.28 & 80.83 \\
    Llama-3.1-8B-Instruct & Google Cloud & 1.56 & 0.99 & 13.11 & 17.73 & 19.63 & 25.92 & 80.31 & 81.85 \\
    Ministral-8B-Instruct-2410 & Google Cloud & 1.14 & 0.87 & 10.16 & \textbf{16.63} & 10.89 & 19.78 & 78.06 & 81.07 \\ \midrule

    GPT-5.2 & Google Workspace & 2.29 & 1.36 & 14.98 & 17.97 & 21.95 & 26.73 & 81.31 & 82.26 \\
    GPT-4.1 & Google Workspace & \textbf{3.08} & 2.16 & 16.09 & 20.94 & 22.64 & 27.33 & \textbf{82.02} & 83.11 \\
    DeepSeek-R1-Distill-Qwen-32B & Google Workspace & 1.43 & 1.56 & 13.04 & 19.19 & 16.55 & 23.55 & 80.24 & 82.04 \\
    Qwen3-32B & Google Workspace & 2.23 & \textbf{2.76} & \textbf{16.52} & 21.28 & 20.81 & 27.08 & 81.97 & \textbf{83.29} \\
    Qwen3-30B-A3B & Google Workspace & 1.88 & 1.68 & 13.72 & 20.14 & 13.72 & 19.46 & 80.16 & 82.38 \\
    DeepSeek-R1-Distill-Llama-8B & Google Workspace & 1.43 & 1.62 & 13.51 & 18.65 & 21.61 & \textbf{27.92} & 80.03 & 81.57 \\
    Llama-3.1-8B-Instruct & Google Workspace & 2.75 & 1.35 & 15.75 & \textbf{21.05} & \textbf{22.93} & 28.93 & 81.17 & 82.74 \\
    Ministral-8B-Instruct-2410 & Google Workspace & 1.31 & 0.85 & 10.75 & 18.37 & 11.95 & 20.13 & 78.93 & 81.63 \\
    \midrule

    GPT-5.2 & GitHub & 0.82 & 0.5 & 10.85 & 13.85 & 17.04 & 20.68 & 78.87 & 79.97 \\
    GPT-4.1 & GitHub & 0.76 & 0.48 & 11.46 & 16.04 & 17.59 & 23.3 & 79.81 & 81.05 \\
    DeepSeek-R1-Distill-Qwen-32B & GitHub & 1.06 & 0.44 & 11.72 & 16.41 & 15.67 & 20.85 & 78.77 & 80.5 \\
    Qwen3-32B & GitHub & 0.88 & 0.37 & \textbf{11.88} & 16.9 & 16.36 & 22.26 & \textbf{80.01} & \textbf{81.92} \\
    Qwen3-30B-A3B & GitHub & \textbf{1.24} & 0.55 & 10.02 & 16.00 & 12.05 & 17.59 & 78.84 & 81.45 \\
    DeepSeek-R1-Distill-Llama-8B & GitHub & 0.71 & 0.27 & 10.26 & 14.39 & \textbf{17.93} & 23.87 & 78.41 & 80.01 \\
    Llama-3.1-8B-Instruct & GitHub & 1.08 & \textbf{0.63} & 11.85 & \textbf{16.87} & 17.28 & \textbf{23.83} & 79.59 & 81.38 \\
    Ministral-8B-Instruct-2410 & GitHub & 0.99 & 0.52 & 9.27 & 16.16 & 10.86 & 18.68 & 78.04 & 80.84 \\
    \midrule

    GPT-5.2 & Atlassian & 0.96 & 0.46 & 11.19 & 14.64 & 16.55 & 21.6 & 79.12 & 80.3 \\
    GPT-4.1 & Atlassian & 1.47 & 0.74 & 12.76 & 17.16 & 17.53 & 23.05 & 79.99 & 81.3 \\
    DeepSeek-R1-Distill-Qwen-32B & Atlassian & \textbf{1.49} & 0.66 & 11.89 & 16.84 & 14.77 & 20.69 & 78.85 & 80.92 \\
    Qwen3-32B & Atlassian & 1.34 & \textbf{0.75} & \textbf{12.81} & \textbf{18.18} & 16.09 & 22.56 & \textbf{80.24} & \textbf{81.9} \\
    Qwen3-30B-A3B & Atlassian & 1.23 & 0.47 & 10.51 & 16.16 & 11.64 & 17.09 & 78.78 & 81.09 \\
    DeepSeek-R1-Distill-Llama-8B & Atlassian & 0.86 & 0.56 & 10.79 & 14.9 & \textbf{17.91} & \textbf{24.22} & 78.98 & 80.56 \\
    Llama-3.1-8B-Instruct & Atlassian & 1.4 & 0.62 & 11.67 & 17.32 & 15.89 & 22.97 & 79.67 & 81.66 \\
    Ministral-8B-Instruct-2410 & Atlassian & 0 & 0.47 & 8.2 & 15.78 & 8.33 & 16.74 & 77.53 & 80.57 \\
\bottomrule
\end{tabular}
\end{table*}

Table \ref{tab:results-root_cause} presents the evaluation performance of the baseline LLMs in identifying root causes under a zero-shot setting across the systems.
Overall, GPT-4.1 demonstrates the strongest and most consistent performance, achieving the highest BERTScore@Top1 on Azure, Google Cloud, Google Workspace, and the highest BLEU-4@Top1 on Azure. Among open-weight models, Qwen3-32B is the strongest, achieving the highest ROUGE-L@Top1 on Azure, Google Workspace, and Atlassian, and the best BERTScore@Top5 on Google Cloud and GitHub. This suggests that Qwen3-32B is a competitive open-weight alternative to GPT-4.1 for RCA, particularly in capturing structural similarity with the ground truth.

DeepSeek-R1-Distill-Llama-8B consistently achieves the highest METEOR@Top1 scores across multiple datasets: Azure, Google Cloud, GitHub and Atlassian indicating strong performance despite its smaller size. In contrast, Ministral-8B-Instruct-2410 consistently underperforms across all datasets and metrics, with BLEU-4@Top1 and BERTScore@Top1 scores consistently lower than other models, suggesting that model scale and architecture play an important role in RCA performance. GitHub presents the most challenging benchmark, with BLEU-4@Top1 scores below 1.3 for all models, while Azure and Google Workspace yield comparatively higher scores, likely reflecting different difficulties across systems. Despite promising results, absolute scores remain modest across all datasets, highlighting the challenges of zero-shot RCA from unstructured incident reports.





\subsubsection{RQ2: \RQTwo}

\begin{table*}
\centering
\caption{Mitigation prediction results. Best results are in bold.}
\label{tab:results-mitigation}
\begin{tabular}{llcccccccc}
\toprule
\multirow{2}{*}{Model} & \multirow{2}{*}{Dataset} & \multicolumn{2}{c}{BLEU-4} & \multicolumn{2}{c}{ROUGE-L} & \multicolumn{2}{c}{METEOR} & \multicolumn{2}{c}{BERTScore} \\
& & Top1 & Top5 & Top1 & Top5 & Top1 & Top5 & Top1 & Top5 \\
\midrule
    GPT-5.2 & Azure & 0 & 0.05 & 9.1 & 12.67 & 15.25 & 20.86 & 77.5 & 78.79 \\
    GPT-4.1 & Azure & \textbf{0.67} & 0.25 & \textbf{10.94} & 15.34 & \textbf{17.22} & \textbf{23.66} & 78.38 & 79.89 \\
    DeepSeek-R1-Distill-Qwen-32B & Azure & 0.48 & 0.28 & 9.21 & 13.23 & 15.47 & 21.26 & 76.85 & 79.33 \\
    Qwen3-32B & Azure & 0 & 0.3 & 9.53 & 13.77 & 16.63 & 22.07 & 78.26 & 79.94 \\
    Qwen3-30B-A3B & Azure & 0 & \textbf{0.35} & 10.71 & \textbf{15.86} & 15.31 & 21.91 & \textbf{79.72} & \textbf{81.19} \\
    DeepSeek-R1-Distill-Llama-8B & Azure & 0.13 & 0.07 & 6.28 & 9.54 & 12.14 & 16.88 & 73.02 & 76.8 \\
    Llama-3.1-8B-Instruct & Azure & 0 & 0.03 & 7.02 & 9.45 & 13.18 & 17.68 & 72.67 & 75.89 \\
    Ministral-8B-Instruct-2410 & Azure & 0.17 & 0.08 & 7.63 & 11.21 & 12.04 & 16.99 & 74.48 & 77.92 \\
    \midrule

    GPT-5.2 & Google Cloud & 0 & 0.06 & 8.19 & 11.31 & 16.32 & 22.11 & 76.5 & 77.73 \\
    GPT-4.1 & Google Cloud & 0.49 & \textbf{0.33} & 10.24 & 14.01 & \textbf{18.57} & \textbf{25.59} & 77.59 & 78.82 \\
    DeepSeek-R1-Distill-Qwen-32B & Google Cloud & 0 & 0.29 & 8.5 & 13 & 16.17 & 22.8 & 75.8 & 78.5 \\
    Qwen3-32B & Google Cloud & 0.38 & 0.3 & 8.67 & 12.93 & 17.94 & 24.15 & 77.65 & 79.37 \\
    Qwen3-30B-A3B & Google Cloud & \textbf{0.64} & 0.3 & \textbf{10.83} & \textbf{16.21} & 18.03 & 25.47 & \textbf{79.66} & \textbf{81.11} \\
    DeepSeek-R1-Distill-Llama-8B & Google Cloud & 0.2 & 0.05 & 6.02 & 9.48 & 13.3 & 19.23 & 73.49 & 76.91 \\
    Llama-3.1-8B-Instruct & Google Cloud & 0 & 0.05 & 5.93 & 8.15 & 13.07 & 17.96 & 72.28 & 75 \\
    Ministral-8B-Instruct-2410 & Google Cloud & 0 & 0.09 & 6.53 & 10 & 13.09 & 18.65 & 74.21 & 77.55 \\
    \midrule

    GPT-5.2 & Google Workspace & 0.43 & 0.15 & 9.41 & 12.35 & 17.22 & 22.52 & 77.45 & 78.43 \\
    GPT-4.1 & Google Workspace & \textbf{0.69} & \textbf{0.32} & 10.84 & 14.77 & \textbf{19.23} & 25.15 & 78.01 & 79.55 \\
    DeepSeek-R1-Distill-Qwen-32B & Google Workspace & 0 & 0.1 & 8.84 & 13.01 & 16.13 & 23.04 & 76.56 & 79.29 \\
    Qwen3-32B & Google Workspace & 0 & 0 & 8.67 & 12.6 & 16.87 & 25.17 & 78.16 & 79.95 \\
    Qwen3-30B-A3B & Google Workspace & 0 & 0 & \textbf{11.45} & \textbf{18.04} & 16.63 & \textbf{26.66} & \textbf{80.07} & \textbf{81.82} \\
    DeepSeek-R1-Distill-Llama-8B & Google Workspace & 0 & 0 & 6.42 & 9.72 & 14.06 & 20.59 & 73.81 & 76.9 \\
    Llama-3.1-8B-Instruct & Google Workspace & 0 & 0.12 & 6.51 & 9.17 & 15.04 & 19.83 & 73.89 & 76.56 \\
    Ministral-8B-Instruct-2410 & Google Workspace & 0.25 & 0.24 & 7.33 & 10.96 & 13.25 & 19.57 & 75.39 & 78.45 \\
    \midrule

    GPT-5.2 & GitHub & 0.53 & 0.26 & 8.06 & 10.98 & 17.04 & 22.21 & 76.56 & 77.72 \\
    GPT-4.1 & GitHub & 0.77 & 0.46 & 8.82 & 12.54 & \textbf{18.09} & \textbf{24.28} & 76.88 & 78.23 \\
    DeepSeek-R1-Distill-Qwen-32B & GitHub & 0.26 & 0.36 & 7.54 & 11.61 & 14.72 & 20.63 & 74.98 & 77.82 \\
    Qwen3-32B & GitHub & 0.69 & \textbf{0.65} & 8.7 & 12.97 & 17.44 & 24.16 & 77.77 & \textbf{79.59} \\
    Qwen3-30B-A3B & GitHub & \textbf{0.82} & 0.4 & \textbf{9.23} & \textbf{14.62} & 15.51 & 23.17 & \textbf{78.8} & 80.49 \\
    DeepSeek-R1-Distill-Llama-8B & GitHub & 0.32 & 0.27 & 5.53 & 8.51 & 13.2 & 18.33 & 73.27 & 76.08 \\
    Llama-3.1-8B-Instruct & GitHub & 0.2 & 0.16 & 4.9 & 6.92 & 11.94 & 16.45 & 71.3 & 74.05 \\
    Ministral-8B-Instruct-2410 & GitHub & 0.23 & 0.23 & 5.6 & 9.79 & 11.44 & 17.08 & 73.58 & 77.13 \\
    \midrule
    
    GPT-5.2 & Atlassian & 0.38 & 0.1 & 9.06 & 12.11 & 17.66 & 22.66 & 77.06 & 78.24 \\
    GPT-4.1 & Atlassian & 0.47 & 0.19 & 11.31 & 15.21 & \textbf{20.12} & \textbf{26.51} & 78.31 & 79.66 \\
    DeepSeek-R1-Distill-Qwen-32B & Atlassian & 0.33 & 0.32 & 9.1 & 13.09 & 16.3 & 22.31 & 75.84 & 78.44 \\
    Qwen3-32B & Atlassian & 0.41 & 0.28 & 10.13 & 14.59 & 17.93 & 25.41 & 78.13 & 80.05 \\
    Qwen3-30B-A3B & Atlassian & \textbf{0.86} & \textbf{0.37} & \textbf{11.39} & \textbf{16.74} & 17 & 24.29 & \textbf{79.76} & \textbf{81.27} \\
    DeepSeek-R1-Distill-Llama-8B & Atlassian & 0.16 & 0.13 & 6.68 & 10.17 & 14.14 & 19.86 & 74.12 & 77.14 \\
    Llama-3.1-8B-Instruct & Atlassian & 0.23 & 0.15 & 6.82 & 9.22 & 13.79 & 18.17 & 72.81 & 75.5 \\
    Ministral-8B-Instruct-2410 & Atlassian & 0.19 & 0.12 & 7.68 & 12.1 & 13.47 & 19.65 & 75.01 & 78.67 \\
\bottomrule
\end{tabular}
\end{table*}

Table \ref{tab:results-mitigation} presents mitigation prediction performance under a zero-shot setting. 
GPT-4.1 achieves the strongest overall performance for METEOR and ROUGE-L, achieving highest METEOR@Top1 on Azure, Google Cloud, Google Workspace, GitHub and Atlassian, and the highest ROUGE-L@Top1 on Azure. Notably, Qwen3-30B-A3B emerges as the strongest open-weight model, outperforming GPT-4.1 on BERTScore across all five datasets and leading on ROUGE-L@Top1 for Google Cloud, Google Workspace, GitHub, and Atlassian. This suggests that the Mixture-of-Experts architecture of Qwen3-30B-A3B may provide advantages for the multi-step reasoning required in mitigation prediction. Qwen3-32B also shows competitive performance on BERTScore@Top5, indicating its ability to generate semantically relevant mitigation steps.
In contrast, smaller models consistently underperform across all datasets. For example, Llama-3.1-8B-Instruct and DeepSeek-R1-Distill-Llama-8B show significantly lower performance across systems and metrics, indicating a significant capability gap relative to larger models. Ministral-8B-Instruct-2410 also consistently underperforms larger models as well.

The consistently low BLEU-4@Top1 scores for mitigation reflect the inherent difficulty of the task. Unlike root cause, mitigation requires system-specific reasoning about  mitigation actions that are not available from incident symptoms. As illustrated in Fig. \ref{fig:sample-prediction}, LLMs tend to generate generic actions rather than precise, context-specific steps, resulting in minimal word overlap with ground-truth. These findings reveal a capability gap in current LLMs in accurately predicting mitigation and motivate future work on domain-adapted approaches.

\subsubsection{RQ3: \RQThree}

To understand the performance of tuned models that learned from historical datasets and how they compare with generic LLMs, we conducted additional experiments using GPT-4.1 and GPT-5.2 with In-Context Learning (ICL).
We focus on ICL as it has been shown to improve RCA performance in prior work \cite{zhang2024automated,chen2024automatic,roy2024exploring,goel2024x}  and is the most widely adopted strategy in existing studies. Unlike fine-tuning, ICL requires no parameter updates or hyper-param tuning, which is known to be sensitive and costly when labeled data is scarce \cite{brown2020language}, while SOTA LLMs can achieve good performance through well-constructed examples alone \cite{min-etal-2022-rethinking}, enabling direct and fair comparison across models.
Following established practices~\cite{zhang2024automated}, we allocated 70\% of each dataset for ICL examples and 30\% for evaluation. For each validation incident, we retrieved the five most relevant incidents from the ICL pool using semantic similarity (with OpenAI's GPT-OSS). Table \ref{tab:icl-results} presents the comparison between zero-shot and ICL settings of these models across the datasets.

\begin{table}[ht]
\centering
\caption{In-Context Learning experiments using GPT-4.1 with Top1 scores (Top5 scores for GPT-4.1 and full results for GPT-5.2 are in Tables VIII \& IX in the appendix \cite{appendixlink}).}
\label{tab:icl-results}
\begin{tabular}{ll@{\hspace{0.5em}}c@{\hspace{0.5em}}c@{\hspace{0.5em}}c@{\hspace{0.5em}}c}
\toprule
\textbf{Task} & \textbf{Dataset} & \textbf{BLEU-4} & \textbf{ROUGE-L} & \textbf{METEOR} & \textbf{BERTScore} \\
\midrule
\multirow{3}{*}{RCA} 
& Azure & 2.4 & 15.59 & 20.93 & 82.42 \\
& GG Cloud & 0 & 12.33 & 17.42 & 81.08 \\
& GitHub & 0 & 12.11 & 15.69 & 80.86 \\
\midrule
\multirow{3}{*}{Mitigation} 
& Azure & 0.98 & 14.07 & 17.34 & 80.78 \\
& GG Cloud & 0 & 12.14 & 20.49 & 79.61 \\
& GitHub & 1.69 & 14.18 & 20.88 & 80.25 \\
\bottomrule
\end{tabular}
\end{table}



\textbf{RCA Insights.} 
ICL yields consistent but modest improvements in semantic similarity across all datasets, while its effect on lexical overlap remains inconsistent. For GPT-4.1, Azure exhibits the clearest gains, especially in BLEU-4@Top1, METEOR@Top1, and BERTScore@Top1. In contrast, Google Cloud and GitHub show divergent behavior between semantic and lexical metrics: BERTScore@Top1 improves marginally on both Google Cloud and GitHub, but BLEU-4@Top1 collapses to 0 in both cases. Google Workspace and Atlassian, meanwhile, achieve the strongest absolute performance under ICL. Taken together, these results indicate that ICL reliably strengthens semantic alignment with the ground truth across all datasets, but its benefit is dataset-dependent.

\textbf{Mitigation Prediction Insights.}  
ICL benefits more consistently with mitigation than with RCA, suggesting that ICL examples give LLMs clearer guidance. For GPT-4.1, Azure improves in ROUGE-L@Top1 and BERTScore@Top1, while Google Cloud shows a similar improvement in METEOR@Top1, even though BLEU-4@Top1 again drops to 0. GitHub sees the biggest improvement overall, especially in BLEU-4@Top1 and BERTScore@Top1, showing ICL works especially well for this dataset. Google Workspace and Atlassian follow the same pattern, reaching especially high in BERTScore@Top1 and METEOR@Top1. These results suggest that the structured, action-oriented nature of mitigation makes it more receptive to ICL than RCA, whose descriptions are more varied and context-specific across incidents.

\textbf{Key Findings.} 
ICL yields more consistent gains for mitigation prediction than for RCA. For RCA, Azure benefits most with BERTScore@Top1 and METEOR@Top1, while Google Cloud and GitHub show mixed results with semantic scores improving slightly but lexical scores declining. For mitigation, gains are more substantial: GitHub improves most with ROUGE-L@Top1 and BERTScore@Top1, while 
Azure improves in ROUGE-L@Top1, suggesting that historical incidents provide more actionable guidance for prediction.
\subsubsection{RQ4: \RQFour} \label{sec:rq4}

To complement quantitative metrics and provide deeper insights into LLM failure modes, we conducted a qualitative analysis of 700 predictions (375 root causes and 375 mitigations) from the top three baselines (GPT-4.1, Qwen3-32B, DeepSeek-R1-Distill-Qwen-32B) (see Tables \ref{tab:rca_qual} and \ref{tab:mitigation_qual}). We randomly sampled 25 incidents from each cloud-based system and manually annotated each prediction using a tailored classification scheme adapted from prior work \cite{roy2024exploring}.

\begin{table}[h]
\centering
\caption{Qualitative analysis of RCA predictions.}
\label{tab:rca_qual}

\textbf{Per-Model Breakdown (Correct Predictions)}\\[2pt]
\begin{tabular}{@{}lcccc@{}}
\toprule
\textbf{Baseline} & \textbf{Total} & \textbf{Prec.} & \textbf{Imp.} & \textbf{Hall.} \\
\midrule
DeepSeek-R1 & 33 & 1 & 32 & 0 \\
Qwen3-32B   & 32 & 2 & 28 & 2 \\
GPT-4.1     & 34 & 4 & 30 & 0 \\
\bottomrule
\end{tabular}

\vspace{10pt}

\textbf{Per-Model Breakdown (Incorrect Predictions)}\\[2pt]
\begin{tabular}{@{}lccccc@{}}
\toprule
\textbf{Baseline} & \textbf{Total} & \textbf{Hall.} & \textbf{Reas. Err.} & \textbf{Insuff. Ev.} & \textbf{Other} \\
\midrule
DeepSeek-R1 & 92 & 26 & 22 & 6 & 38 \\
Qwen3-32B   & 93 & 22 & 31 & 3 & 37 \\
GPT-4.1     & 91 & 36 & 20 & 7 & 28 \\
\bottomrule
\end{tabular}

\vspace{4pt}
\raggedright
\footnotesize\textit{Prec.: Precise, Imp.: Imprecise, Hall.: Hallucination, Reas. Err.: Reasoning Error, Insuff. Ev.: Insufficient Evidence, DeepSeek-R1: DeepSeek-R1-Distill-Qwen-32B.}

\end{table}

\begin{table}[h]
\centering
\caption{Qualitative analysis of mitigation predictions.}
\label{tab:mitigation_qual}

\textbf{Per-Model Breakdown (Correct Predictions)}\\[2pt]
\begin{tabular}{@{}lcccc@{}}
\toprule
\textbf{Baseline} & \textbf{Total} & \textbf{Prec.} & \textbf{Imp.} & \textbf{Hall.} \\
\midrule
DeepSeek-R1 & 20 & 1 & 19 & 0 \\
Qwen3-32B   & 28 & 2 & 22 & 4 \\
GPT-4.1     & 49 & 5 & 41 & 3 \\
\bottomrule
\end{tabular}

\vspace{10pt}

\textbf{Per-Model Breakdown (Incorrect Predictions)}\\[2pt]
\begin{tabular}{@{}lccccc@{}}
\toprule
\textbf{Baseline} & \textbf{Total} & \textbf{Hall.} & \textbf{Reas. Err.} & \textbf{Insuff. Ev.} & \textbf{Other} \\
\midrule
DeepSeek-R1 & 105 & 57 & 11 & 5 & 32 \\
Qwen3-32B   & 97  & 69 & 9  & 1 & 18 \\
GPT-4.1     & 76  & 51 & 9  & 0 & 16 \\
\bottomrule
\end{tabular}

\vspace{4pt}
\raggedright
\footnotesize\textit{Prec.: Precise, Imp.: Imprecise, Hall.: Hallucination, Reas. Err.: Reasoning Error, Insuff. Ev.: Insufficient Evidence, DeepSeek-R1: DeepSeek-R1-Distill-Qwen-32B.}

\end{table}
 

\textbf{Classification Scheme.} For each prediction-reference pair, annotators 
assessed semantic alignment with the reference, then classified each prediction into a category reflecting its correctness or error type
\cite{roy2024exploring}. Correct predictions were classified as: (1) \textit{Precise} (precisely matching the reference), (2) \textit{Imprecise} (semantically matching but missing details), or (3) \textit{Hallucination} (semantically correct but containing factual errors). Incorrect predictions were classified as: (1) \textit{Hallucination} (factual errors in content), (2) \textit{Reasoning Error} (logical errors in reasoning), (3) \textit{Insufficient Evidence} (refraining from making a prediction), or (4) \textit{Other} (unknown error). This taxonomy enables the identification of failure modes and ensures a context-aware assessment of predictions.

\textbf{RCA Insights.} LLMs achieve modest RCA correctness (26.4\%). While correct predictions are often semantically aligned, they often lack technical precision. Errors are dominated by hallucinations and faulty reasoning, where LLMs predict plausible but incorrect information or misinfer causal chains. Model-specific patterns show trade-offs: Qwen3-32B shows the highest reasoning error rate, likely from its reasoning chain, whereas DeepSeek-R1 balances both error types. GPT-4.1 leans toward hallucinations, suggesting overconfident pattern generation misaligned with the ground truth. These failures likely stem from limited evidence, plausible-sounding predictions, and over-generalized root causes.

\textbf{Mitigation Prediction Insights.} Performance is weaker for mitigation, with correctness of only 25.87\%. Hallucination is the dominant failure mode, often triggered by insufficient summaries that lead models to predict unrelated steps. Qwen3-32B shows the highest hallucination rate but fewer reasoning errors, while GPT-4.1 reduces hallucinations at the cost of more reasoning errors; DeepSeek-R1 falls between these extremes. These limitations reflect retrieval constraints, pattern-matching biases, and a tendency to generate generic rather than specific technical actions.

\begin{figure}[ht]
\fbox{\begin{minipage}{0.98\columnwidth}\small
\textbf{Title:} Incident affecting Admin Console

\textbf{Summary:}
The incident report documented a 5 hour and 30-minute period on Monday, 18 July 2022, during which Google Cloud and Google Workspace Support customers were unable to attach files to new and existing support cases via the Cloud Console and Admin Console. This affected 100\% of users for both services, with support case creation remaining unaffected. The incident impacted file uploads specifically, while support case creation functionality remained intact.

\textbf{Reference root cause:}
File upload service became inaccessible due to a configuration change in the Global Service Load Balancer, causing intermittent failures for users uploading files.

\textbf{Reference mitigation:}
The incident was mitigated by disabling the file attachment feature and rolling back the GSLB configuration change.

\textbf{Predicted root cause:}
The root cause was a failure in the file upload functionality within the Admin Console, likely due to an issue with the backend service, API, or database handling file uploads, preventing users from successfully attaching files to support cases.

\textbf{Predicted mitigation:}
The incident was resolved by identifying the root cause, implementing a fix to restore file upload functionality, communicating updates to users, and conducting a post-mortem analysis to improve future response processes
\end{minipage}}
\centering
\caption{A sample of failed prediction but with high scores.}
\label{fig:sample-prediction}
\end{figure}

\textbf{Key Findings.} Our qualitative analysis shows that current LLMs face fundamental challenges in incident report-based RCA and mitigation prediction, with most errors stemming from hallucinations and reasoning failures due to contextual sparsity and limited external evidence. While models can capture high-level patterns, they often omit critical technical details, rely on generic or vague language, fabricate plausible-sounding causes, and repeat surface-level details from the input (see Fig. \ref{fig:sample-prediction}).
In this case, the model correctly identifies the affected component (file upload) but incorrectly attributes the root cause to a generic ``backend service, API, or database" issue rather than the actual configuration change in the Global Service Load Balancer. Despite this error, the prediction still receives high scores for two reasons. First, BLEU-4 and ROUGE-L reward high-frequency terms shared between prediction and ground truth, such as ``file upload," ``users", and phrases like ``The incident was mitigated/resolved by" regardless of causal correctness. Second, root cause descriptions often follow an "A due to B" pattern, where A represents the observed consequence, and B represents the underlying root cause. Models tend to reproduce A, which is usually mentioned in the incident summary, while failing to identify B. Thus, predictions that accurately describe the observed consequences but misidentify the underlying root causes may still receive high scores, reflecting a limitation of automatic evaluation metrics in reasoning-intensive tasks.

\section{Threats to Validity} \label{sec:threats}

Despite substantial manual effort to ensure the accuracy and reliability of the ground truth annotations, the labeling process is inherently constrained by the lack of direct engagement with the cloud-based system owners. As the label annotators are not the original owners of the incident reports, our understanding of certain cases may be limited, and the assigned labels for root causes and mitigation steps may reflect subjective interpretations given the available information. This can introduce potential bias or misalignment with the actual intent or context of incident reports. This limitation is mitigated by involving two independent annotators in the labeling process and conducting consensus discussions to resolve discrepancies and agree on the final label for each report. To increase the internal validity of our evaluation, we generate 10 predictions per incident to capture output diversity. We further assess all methods using four metrics and qualitative analysis.

Besides, using public post-incident reports introduces specific limitations. Variability and incompleteness are inherent characteristics of real-world incident reports \cite{ahmed2023recommending}, rather than a limitation unique to our work, as reports differ in detail, terminology, and technical depth across systems. As a result, ground-truth root causes and mitigation actions reflect what engineers choose to document under operational and disclosure constraints and may not capture all causal factors. Nonetheless, using authentic reports is a practical choice that preserves realism by exposing LLMs to the variability, ambiguity, and incompleteness in real-world scenarios.

Third, lexical and semantic metrics measure textual overlap and similarity with the ground truth, but do not necessarily capture the correctness of RCA and mitigation predictions, as demonstrated in our qualitative analysis (see Section~\ref{sec:rq4}). Nevertheless, we adopt these metrics as they are widely used in prior work \cite{ahmed2023recommending,zhang2024automated,roy2024exploring,saha2022mining,goel2024x,wang2024rcagent}, ensuring our results are comparable with those of existing studies. We further complement these metrics with a qualitative analysis to examine whether the predictions correctly identify root causes and recommend appropriate mitigation actions. In future work, we will investigate alternative evaluation approaches such as LLM-as-a-judge \cite{zheng2023judging} to better assess prediction correctness.

Fourth, as the incident reports are crawled from public web pages, there could be a chance that some evaluated LLMs may have seen the same or similar reports during pretraining. However, the relatively low performance of the evaluated LLMs suggests that their predictions are unlikely to be primarily driven by memorization of the original incident reports. Nevertheless, as with other publicly available benchmarks, we cannot entirely rule out potential data contamination, as the pretraining data of the LLMs are not fully disclosed.

Lastly, our evaluation covers a fixed set of LLMs. These models were selected from among the top-performing and most widely adopted models on SWE-bench at the time of this study. As new models are frequently released, in future work, we will extend the evaluation to stronger LLMs through standard inference APIs.

\section{Conclusions}
In this work, we introduce \DatasetName, the first publicly available benchmark for incident report-based RCA and mitigation prediction. \DatasetName addresses a critical gap in the literature by enabling reproducible research in these tasks. It includes a dataset of incident reports curated from five leading cloud-based systems, with each report containing detailed metadata and ground truth labels for root causes and mitigation strategies. In addition, we provide a comprehensive evaluation environment consisting of recent LLM-based methods, further facilitating research in these areas. Finally, we conduct a comprehensive evaluation that highlights both the strengths and limitations of current LLM-based methods in RCA and mitigation prediction.
For future work, we plan to explore strategies such as fine-tuning, retrieval-augmented generation (RAG), agent-based reasoning, and LLM-as-a-judge to advance LLM performance for our benchmark.



\printbibliography

\end{document}